# Item Recommendation Using User Feedback Data and Item Profile

Debashish Roy[1, a)], Rajarshi Roy Chowdhury[2, 3], Abdullah Bin Nasser[4], Afdhal Azmi[5] and Marzieh Babaeianjelodar[6]

[1] *Information Technology Management, Ryerson University, 350 Victoria Street, Toronto, ON M5B 2K3, Canada*
[2] *Faculty of Integrated Technologies, Universiti Brunei Darussalam, Jalan Tungku Link, Gadong BE1410, Brunei Darussalam*
[3] *Department of Computer Science and Engineering, Sylhet International University, Shamimabad Road, Sylhet 3100, Bangladesh*
[4] *School of Technology and Innovation, University of Vaasa, FI-65200 Vaasa, Finland*
[5] *Department of Electrical and Computer Engineering, Faculty of Engineering, Universitas Syiah Kuala, Banda Aceh, Indonesia*
[6] *Department of Computer Science, Clarkson University, New York, USA*

Corresponding author: [a)] *debashish.roy@ryerson.ca*

**ABSTRACT.** Matrix factorization (MS) is a collaborative filtering (CF) based approach, which is widely used for recommendation systems (RS). In this research work, we deal with the content recommendation problem for users in a content management system (CMS) based on users' feedback data. The CMS is applied for publishing and pushing curated content to the employees of a company or an organization. Here, we have used the users' feedback data and content data to solve the content recommendation problem. We prepare individual user-profiles and then generate recommendation results based on different categories, including Direct Interaction, Social Share and Reading Statistics, of user's feedback data. Subsequently, we analyze the effect of the different categories on the recommendation results. The results have shown that different categories of feedback data have different impacts on recommendation accuracy. The best performance achieves if we include all types of data for the recommendation task. We also incorporate content similarity as a regularization term into an MF model for designing a hybrid model. Experimental results have shown that the proposed hybrid model demonstrates better performance compared with the traditional MF-based models.

## INTRODUCTION

The problem of information overload has become prominent owing to the rapid growth in the amount of available digital information and increasing the number of visitors to different web resources, such as music, movies and e-book. Due to massive amount (or volume) of information, it is challenging for the users to find items of interest in a short time. Finding users preferable items, recommender systems (RS) are used widely [1–3]. Recommender systems are information filtering techniques that recommend items to different users according to their preferences, interests, or observed behaviour on items. Depending on a user profile, RS can determine whether a particular user likes an item or not. Hence, various approaches have been developed by researchers for building recommender systems, including collaborative filtering (CF), content-based filtering, and hybrid filtering [4]. Nowadays, RS have been applied in different research fields, such as Internet of Things (IoT) [5–7], social networks

[8], audio and video contents [9–11], news [12], for efficient information management in terms of time and cost [13].

Most of the CF-based RS [14, 15] use a rating matrix to recommend items or products. Rating matrix is generated from the value of implicit or explicit ratings (or scores) given by the users on different items. Rating data may be retrieved from different content provider websites, which can be used as explicit feedback, whilst user interaction data, such as likes, comments, etc., can be used as implicit feedback. However, most of the CF-based recommender systems do not include text information of the items or products in their rating matrix. The CF-based recommenders suffer a lot when the rating matrix is sparse. To improve the recommendation accuracy, we can include text information into the CF-models. In this work, we generate a user-item rating matrix based on the feedback record from different users and include content information. The main objectives for this work are:

- To study users interactions and utilizing their feedback data stored in a database to generate users-profile. Here we analyze different types of interactions that a user makes on their posts. We calculate implicit feedback scores based on the users' historical interaction records on a post and then prepare individual user profiles using their interaction data on different posts.
- Utilizing the CF method for content recommendation, we utilize users' feedback score as the rating data and a MF model on the rating matrix for content recommendation.
- Finding similarity scores between different posts and incorporating these as an extra regularization term into the MF model.
- Analyzing the effect of different categories of feedback data on the recommendation results.

The remaining of this paper is organized as follows. Section 2 presents related works in the field of recommender systems. The proposed recommender system architecture and matrix factorization scheme are given in Section 3. Section 4 describes the experimental details used to evaluate the proposed model performances. Finally, Section 5 summarizes our findings, highlights the contributions, and points out future research directions.

## RELATED WORK

Researchers in different domains have utilized recommender systems for improving decision making and the quality of services in the context of users and service providers. For a personalized recommendation, a clustering-based approach is proposed in reference [16] to model a user profile, whilst the authors selected various attributes, including interest, profession, gender, country, languages, postal code, age, etc., for constructing a user model. Using these attributes, they created communities of user-profiles by gathering similar profiles, which constitutes the clustering phase of their approach. Similarly, both explicit and implicit ratings have been used in reference [17] to model a user profile. For explicit profiling, the authors collect explicit opinions of the users, which are given by the users about an item after consuming its content. On the other hand, for implicit profiling, the authors have used duration, e.g. browsing duration, which a user spends for the content of an item. For instance, a longer consumption time for a user on an item, such as a video, implies their preference for that particular item. Then using a weighted combination of these ratings, they created the user model that the recommender engine would use to predict unseen items for a user. To extract signals about individual preferences for a recommendation, users personality traits have been used in references [18, 19]. The researchers assessed user personality along with user ratings to deliver a recommendation.

Recommender systems are used widely to recommend web services, e.g. e-commerce. To generate a user profile using explicit feedback is hard, because users' ratings on services seldomly report. In reference [20], implicit feedback data are used to build a user profile for a web-service recommendation, whilst the authors used user-service interaction records, including duration and frequency of uses, for building a user profile. In prior works, the researchers have utilized different approaches to build user profiles for recommender systems. However, in the proposed work, we utilize users' explicit, implicit, and social share interactions to build a user profile. Later, a weighted combination of these three types of interactions evaluates for calculating a combined feedback score to generate users-profile. Additionally, for generating users' profiles weighted scores for the individual types of feedback have been used.

# METHODOLOGY

This section describes the proposed methods for content recommendations utilizing user feedback data. In this analysis, a company's content management system (CMS) dataset has been used for evaluating the proposed model performance. The CMS dataset consists of employee interactions on different posts, including direct posts, sharing a post either locally or on social network platforms. This research work has explored two recommendation methods: collaborative filtering (CF) approach (using only the feedback data) and hybrid schemes (using CF and post-similarity scores).

## Workflow of the Proposed Approaches

To accomplish this research goal, we have taken the following steps: i) we generate individual user profiles based on their interactions with the posts, whilst interactions are made by users with their posts, including like, share, comment, tweet, and reply. After extracting user interactions, we classify these into three major categories based on the nature of interactions. The first category is Direct Interaction, which means that user/post-interaction happened in the CMS itself. The second category is Social Share, which means users share posts from the CMS system, to their social networking websites. The third category is Reading Statistics, which refers to the feedback score a user gives to the particular content by taking a specific action, such as reading progress (reading action). In the CMS, all the interactions have been assigned individual weights. We use those weights and calculate the user feedback score for each user. Furthermore, we compare results from each category with an overall feedback score to evaluate which category reflects the user behaviour the most; ii) we generate content recommendations using a CF algorithm, whilst we utilize a CF approach and users implicit feedback scores as rating data. We also use an MF algorithm for the collaborative filtering model to generate recommendations for individual users; iii) we find similarity scores between different posts, whilst these scores are added to the MF model as a regularization term; iv) finally, we evaluate recommendation results received from a) MF based on ratings and b) MF with regularization term based on post similarities. An abstract view of the proposed recommendation approach is depicted in Figure 1.

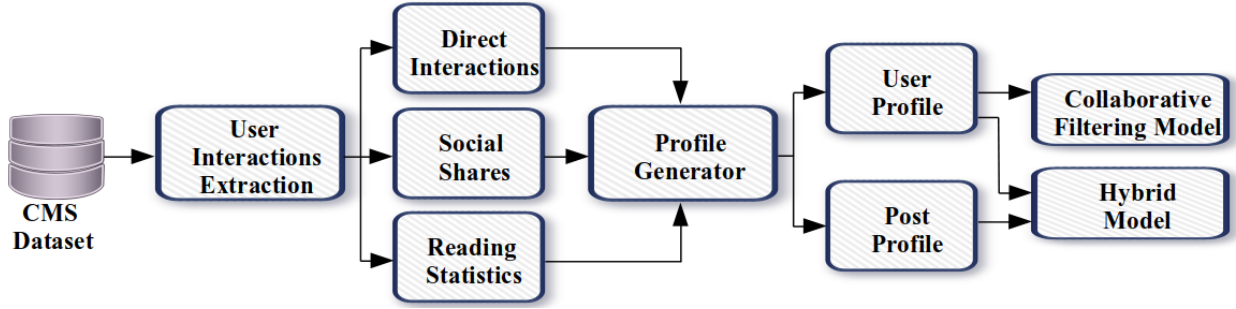

**FIGURE 1.** Workflow of the proposed model.

## Matrix Factorization

The matrix factorization (MF) [21, 22] is used extensively in recommender systems. It uses various known rating scales for generating the latent preferences of users and latent features of the items. To predict the unknown ratings, we multiply the user latent vectors and item latent vectors. The process of MF starts with a user-item rating matrix $R$. Here, the size of matrix $R$ is $m \times n$, where $m$ denotes the total number of users and $n$ denotes the total number of items. MF decomposes the rating matrix $R$ into two low-rank latent feature matrices $P$ for users and $Q$ for items, whilst the size of matrix $P$ is $m \times d$ and the size of matrix $Q$ is $n \times d$, whilst $d$ is the rank of the matrices and defines dimension for the latent features [23]. If $\hat{R}$ is a matrix of predicted ratings, then matrix factorization approximates $\hat{R}$ in such a way that $\hat{R} = PQ^T$. Afterwards, we use the matrices $P$ and $Q$ for predicting the ratings on different items from any users. Predicting a rating scale from user $u$ on item $i$, an inner product between $P_u$ and $Q_i$ is performed. To apply MF, different optimization techniques have been proposed. For decomposing a sparse rating matrix, the following objective function (Equation 1) is used by the traditional matrix factorization methods [24]:

$$L = \min_{P,Q} \frac{1}{2} \sum_{(u,i) \in C} (R_{u,i} - P_u Q_i^T)^2 + \frac{\lambda}{2} (\|P\|_F^2 + \|Q\|_F^2) \text{ -------- (1)}$$

where *C* indicates a set of (user, item) pairs of known ratings, *T* defines transpose of *Q* matrix and $\|.\|F$ represents a matrix norm (*F* means Frobenius norm). To avoid overfitting problem, two regularization terms on the sizes of *P* and *Q* are added as constraints and $\lambda$ is used as a regularization parameter. In this proposed work, to complete MF, we have used the rating data and the objective function *L* defined in Equation 1. However, the objective function *L* in Equation 1 is only based on users' ratings, it does not include the content information in the factorization model. To do so, we include the post similarity score into the model. We define $S_{jn}$ as the similarity coefficient between two posts j and n, which satisfies: i) $S_{jn} \in [0,1]$; ii) $S_{jn} = S_{nj}$; and iii) the larger $S_{jn}$ means, the more similarity between the posts. With the similarity coefficient, the similarity regularization is to minimize (*min*) the following term:

$$\min \frac{\alpha}{2} \sum_{j=1}^{N} \sum_{n=1}^{N} (S_{jn} - Q_j^T Q_n)^2 \text{ -------- (2)}$$

To add the impact of content profiles to the basic factorization model, we add a regularization term defined in Equation 2, where *N* defines the total number of items and $\alpha$ represents a regularization parameter, to the objective function *L* (Equation 1). Hence, we get the following new objective function *L^* (Equation 3):

$$\hat{L} = \min_{P,Q} \frac{1}{2} \sum_{(u,i) \in C} (R_{u,i} - P_u Q_i^T)^2 + \frac{\lambda}{2} (\|P\|_F^2 + \|Q\|_F^2) + \frac{\alpha}{2} \sum_{j=1}^{N} \sum_{n=1}^{N} (S_{jn} - Q_j^T Q_n)^2 \text{ -------- (3)}$$

In this research, we design the hybrid recommendation model based on the updated objective function *L^* and use this model to complete MF based on both rating scale data and content similarity score. Additionally, we employ this model to multiple user profiles.

## Posts Similarity Calculation

Finding the similarity score between posts, we need to build content profiles. For modelling the post profiles a potential solution is used, such as bag-of-words (BoW), i.e. term frequency-inverse document frequency (TF-IDF) [25]. In this experiment, we generate content profiles using the TF-IDF. Hence, finding similarities between different posts, we utilize the cosine similarity algorithm. However, in Section 3, we have been explained the proposed methodology of this research work. In the following section (Section 4), we discuss the experimental details that have been performed for evaluating the proposed approach, including details of the overall system.

## EXPERIMENTS

In this section, we explore model implementation, experimental design, and the experiments that we have performed to validate the performance of our proposed recommendation systems.

## Model Implementation

The RS model was implemented on a Windows system (version 10) with Intel Core i7 processor, 16 GB RAM and 2.0 GHz clock-speed. Python (version 2.7) is used as the programming language for the implementation of the proposed model. Since it has most of the libraries used for machine learning (ML) and data mining techniques. The CMS database is recorded in my structured query language (MySQL) server; hence, MySQL connector is used in Python code for extracting data required for this research.

## Dataset

The dataset used in this experiment is a company's CMS dataset. It has two significant entities: posts and users, which consists of 250 users, and 6,900 posts as well as user/post interactions. For each post, we extract the post ID and contents of the post. Similarly, for each user, we extract user ID and their interaction data, including direct share, direct impression, direct reshare, direct like, direct comment, direct clickthrough, Twitter share/reshare, Facebook share/reshare, LinkedIn share/reshare, reading progress, etc. We have been using all these interactions for generating user profiles. First, we categorize all interactions into three categories: Direct Interaction, Social Share, and Reading Statistics, whilst Tables 1, 2, and 3 show the statistical information about each category, respectively. Then, we generate an overall feedback score considering these three categories. We also calculate the feedback score by considering each category separately, to identify which category has a greater impact on the recommendations system.

**TABLE 1.** Statistical information about Direct Interaction category.

| Description | Item counts |
| --- | --- |
| Total number of direct interactions | 20,868 |
| Total number of users with direct interactions | 150 |

**TABLE 2.** Statistical information about Social Share category.

| Description | Item Counts |
| --- | --- |
| Total number of social shares | 28,363 |
| Total number of users with social shares | 165 |

**TABLE 3.** Statistical information about Reading Statistics category.

| Description | Number of Interactions |
| --- | --- |
| Total number of posts with reading statistics | 10,985 |
| Total number of users with reading statistics | 134 |

## Evaluation Results: RMSE, MAE, and F1-Score

We have used both basic MF and hybrid MF, to generate recommendations based on different types of feedback matrices. Root mean squared error (RMSE) and mean absolute error (MAE) values are calculated for the generated recommendation results. And, for the same result sets, we also estimated F1-scores for the top 10 recommendations. Figures 2 and 3 show that all interaction-based recommendation systems generate the smallest RMSE and MAE values for both factorization methods using the datasets, as presented in Tables 1, 2 and 3.

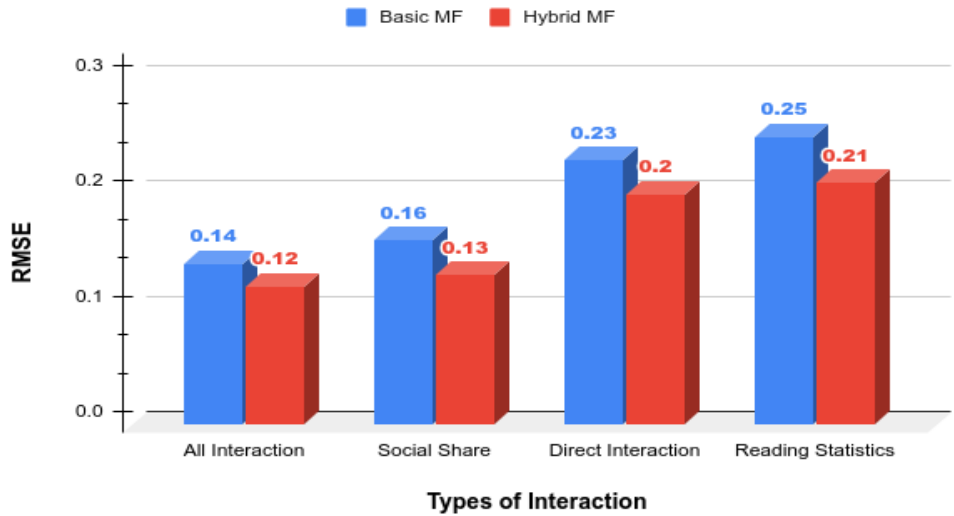

**FIGURE 2.** Evaluation of the model using RMSE.

From Figure 4, we observed that all interaction-based recommendations generate the highest F1-score for both models: basic MF and hybrid MF, using all the available datasets listed in Tables 1, 2 and 3. We also notice that out of these three types of feedback categories, e.g. social, direct, and reading, social share-based recommendation performed the best, and the worst-performing one is the reading statistics-based recommendation. However, in all cases, the hybrid model outperformed the traditional model, whilst performances of the social share are very close to all interaction performances.

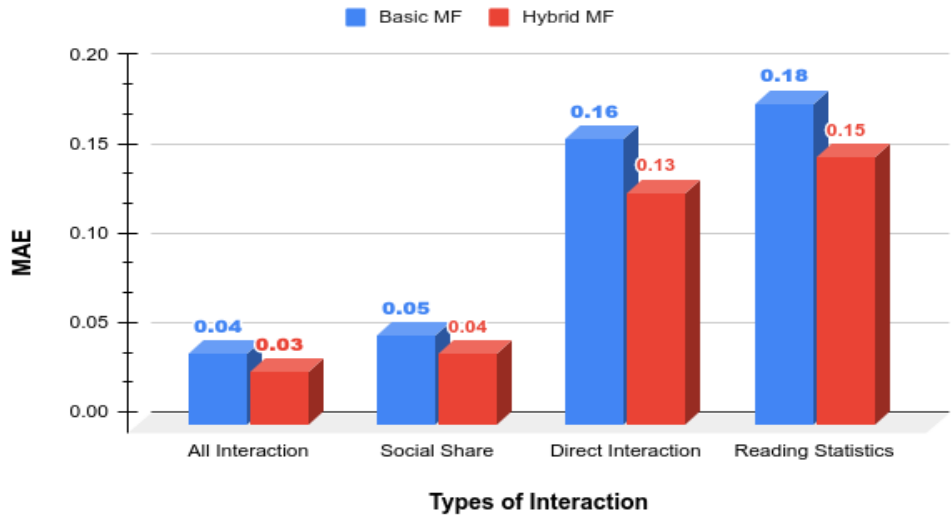

**FIGURE 3.** Evaluation of the model using MAE.

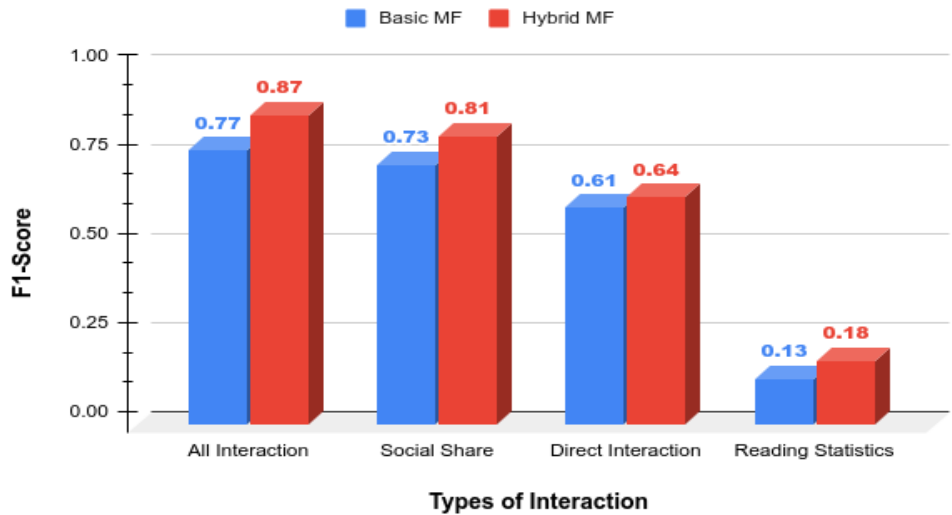

**FIGURE 4.** Evaluation using F1-Score.

## Discussion

We have four categories of user-profiles: All Interaction, Direct Interaction, Social Share, and Reading Statistics. The result shows that the All-Interaction feedback score profile is performed the best among all four profiles. If we consider three profile categories: Social Share, Direct Interaction, and Reading Statistics, then the Social Share category performs much better than the other two types. We get the second-best performance from the Direct Interaction profile, and the last one is the Reading Statistics profile. The reason behind this result is that there are different types of social shares in this category. In other words, we can say in this category; we have more types of user/post interactions than the other categories, and data largely sharing by users on their social networking websites. As a comparison, the Reading Statistics profile has only two types of interactions. Therefore, the Social Share profile has more interaction data. It works better than the Direct Interaction profile and Reading Statistics profile. While comparing the Social Share profile with the All-Interaction feedback score profile, the All-Interaction feedback score profile works better than the Social Share profile. Though, their differences are very small. However, when efficiency is a concern, we may use social feedback scores only, because it requires minimal data processing time. Finally, we compare all the evaluation metrics based on the two methods of matrix factorization: basic MF and hybrid MF. The results show that the hybrid MF performs better than the basic MF for all four types of user-profiles.

## CONCLUSION

This research evaluates the effectiveness of generating content profiles by utilizing different types of user feedback data to provide content recommendations. Overall, the evaluation result shows that combining TF-IDF content profiles with all interaction feedback user profile, i.e. the hybrid model generates the best recommendations for this CMS dataset. Moreover, different types of user interaction data may have different impacts on recommendation performance. Therefore, exploring and investigating the optimal way of using and combining them in a recommender system is necessary. In future work, we want to compare automatically generated user interaction weights with those weights that are predefined in a dataset. Additionally, we would like to apply a regression model or a classification algorithm for finding optimal weights based on user interactions.

## ACKNOWLEDGEMENTS

The authors are profoundly grateful to the Ted Rogers School of Information Technology Management, Ryerson University, for supporting this research work.

## REFERENCES

1. C. C. Aggarwal, Recommender systems. Springer International Publishing Switzerland, (2016).
2. J. Chen, H. Dong, X. Wang, F. Feng, M. Wang, and X. He, Bias and Debias in Recommender System: A Survey and Future Directions, (2020) 1–20.
3. B. Walek and V. Fojtik, A hybrid recommender system for recommending relevant movies using an expert system, Expert Syst. Appl. 158 (2020). doi: 10.1016/j.eswa.2020.113452.
4. V. Vekariya and G. R. Kulkarni, Hybrid recommender systems: Survey and experiments, 2012 2nd Int. Conf. Digit. Inf. Commun. Technol. its Appl. DICTAP 2012, (2012) 469–473. doi: 10.1109/DICTAP.2012.6215409.
5. M. Altulyan, L. Yao, X. Wang, C. Huang, S. S. Kanhere, and Q. Z. Sheng, Recommender Systems for the Internet of Things: A Survey, (2020).
6. R. R. Chowdhury, S. Aneja, N. Aneja, and E. Abas, Network Traffic Analysis based IoT Device Identification, in ACM International Conference Proceeding Series, 06 (2020) 79–89. doi: 10.1145/3421537.3421545.
7. R. R. Chowdhury, S. Aneja, N. Aneja, and P. E. Abas, Packet-level and IEEE 802.11 MAC frame-level Network Traffic Traces Data of the D-Link IoT devices, Data Br. 37 (2021). doi: 10.1016/j.dib.2021.107208.
8. S. Deng, L. Huang, G. Xu, X. Wu, and Z. Wu, On Deep Learning for Trust-Aware Recommendations in Social Networks, IEEE Trans. Neural Networks Learn. Syst. 28 (2017) 1164–1177. doi: 10.1109/TNNLS.2016.2514368.
9. D. C. G. Putri, J. S. Leu, and P. Seda, Design of an unsupervised machine learning-based movie recommender system, Symmetry (Basel). 12 (2020) 1–27. doi: 10.3390/sym12020185.
10. D. Roy and C. Ding, Movie Recommendation using YouTube Movie Trailer Data as the Side Information, Proc. 2020 IEEE/ACM Int. Conf. Adv. Soc. Networks Anal. Mining, ASONAM 2020, (2020) 275–279. doi: 10.1109/ASONAM49781.2020.9381349.
11. D. Roy and C. Ding, Multi-source based movie recommendation with ratings and the side information, Soc. Netw. Anal. Min. 11 (2021). doi: 10.1007/S13278-021-00785-5.
12. S. Cao, N. Yang, and Z. Liu, Online news recommender based on stacked auto-encoder, Proc. - 16th IEEE/ACIS Int. Conf. Comput. Inf. Sci. ICIS 2017, (2017) 721–726. doi: 10.1109/ICIS.2017.7960088.
13. F. O. Isinkaye, Y. O. Folajimi, and B. A. Ojokoh, Recommendation systems: Principles, methods and evaluation, Egypt. Informatics J. 16 (2015) 261–273. doi: 10.1016/j.eij.2015.06.005.

14. J. Wei, J. He, K. Chen, Y. Zhou, and Z. Tang, Collaborative filtering and deep learning based recommendation system for cold start items, Expert Syst. Appl. 69 (2017) 1339–1351. doi: 10.1016/j.eswa.2016.09.040.
15. J. Bobadilla, S. Alonso, and A. Hernando, Deep learning architecture for collaborative filtering recommender systems, Appl. Sci. 10 (2020). doi: 10.3390/app10072441.
16. S. Ouaftouh, A. Zellou, and A. Idri, UPCAR: User profile clustering based approach for recommendation, ACM Int. Conf. Proceeding Ser. (2017) 17–21. doi: 10.1145/3175536.3175568.
17. D. Weiß, J. Scheuerer, M. Wenleder, A. Erk, M. Gülbahar, and C. Linnhoff-Popien, A user profile-based personalization system for digital multimedia content, Proc. - 3rd Int. Conf. Digit. Interact. Media Entertain. Arts, DIMEA 2008, (2008) 281–288. doi: 10.1145/1413634.1413687.
18. R. P. Karumur, T. T. Nguyen, and J. A. Konstan, Personality, User Preferences and Behavior in Recommender systems," Inf. Syst. Front., 20 (2018) 1241–1265. doi: 10.1007/s10796-017-9800-0.
19. T. T. Nguyen, F. Maxwell Harper, L. Terveen, and J. A. Konstan, User Personality and User Satisfaction with Recommender Systems, Inf. Syst. Front., 20 (2018) 1173–1189. doi: 10.1007/s10796-017-9782-y.
20. G. Tian, J. Wang, K. He, C. Sun, and Y. Tian, "Integrating implicit feedbacks for time-aware web service recommendations," Inf. Syst. Front., vol. 19, no. 1, pp. 75–89, 2017, doi: 10.1007/s10796-015-9590-1.
21. P. Lops, M. De Gemmis, and G. Semeraro, Recommender Systems Handbook, (2011).
22. D. Roy, Recommending Curated Content Using Implicit Feedback, Asian J. Res. Comput. Sci. 5 (2020) 10–16. doi: 10.9734/ajrcos/2020/v5i230130.
23. S. Deerwester, S. T. Dumais, G. W. Furnas, T. K. Landauer, and R. Harshman, Indexing by latent semantic analysis," J. Am. Soc. Inf. Sci. 41 (1990) 391–407. doi: 10.1002/(SICI)1097-4571(199009)41:6<391::AID-ASI1>3.0.CO;2-9.
24. F. Li, G. Xu, and L. Cao, Two-level matrix factorization for recommender systems, Neural Comput. Appl. 27 (2016) 2267–2278. doi: 10.1007/s00521-015-2060-3.
25. G. Zhao, Y. Liu, W. Zhang, and Y. Wang, TFIDF based feature words extraction and topic modeling for short text, ACM Int. Conf. Proceeding Ser. (2018) 188–191. doi: 10.1145/3180374.3181354.